%% file: JemEUSO-OffLine_MASTER.tex
\documentclass[a4paper,11pt]{article}
\usepackage{jinstpub} 
\usepackage{lineno}
\usepackage{xcolor}
\usepackage{units}
\usepackage{url}
\usepackage{caption}
\usepackage{subcaption}

\newcommand{\Offline}{\mbox{$\overline{\rm Off}$\hspace{.05em}\raisebox{.3ex}{$\underline{\rm line}\enspace$}}}
\DeclareRobustCommand{\OfflineT}{\mbox{$\overline{\rm {\bf Off}}$\hspace{.05em}\raisebox{.3ex}{$\underline{\rm {\bf line}}\,$}}}

\title{EUSO-$\OfflineT$: A Comprehensive Simulation and Analysis Framework}

\collaboration{\includegraphics[height=17mm]{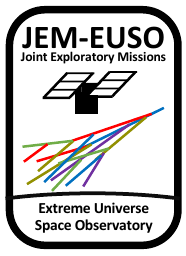}\\[6pt] JEM-EUSO collaboration}

\author[ff]{S.~Abe,}
\author[ld]{J.H.~Adams Jr.,}
\author[cb]{D.~Allard,}
\author[ld]{P.~Alldredge,}
\author[ep]{R.~Aloisio,}
\author[le]{L.~Anchordoqui,}
\author[ed,eh]{A.~Anzalone,}
\author[ek,el]{E.~Arnone,}
\author[cb]{B.~Baret,}
\author[el,ek,em]{D.~Barghini,}
% ...

% ...
\affiliation[cb]{Universit\'e Paris Cit\'e, CNRS, AstroParticule et Cosmologie, F-75013 Paris, France}
% ...
\affiliation[ed]{Istituto Nazionale di Fisica Nucleare - Sezione di Catania, Italy}
% ...
\affiliation[eh]{INAF - Istituto di Astrofisica Spaziale e Fisica Cosmica di Palermo, Italy}
% ...
\affiliation[ek]{Istituto Nazionale di Fisica Nucleare - Sezione di Torino, Italy}
\affiliation[el]{Dipartimento di Fisica, Universit\`a di Torino, Italy}
\affiliation[em]{Osservatorio Astrofisico di Torino, Istituto Nazionale di Astrofisica, Italy}
% ...
\affiliation[ep]{Gran Sasso Science Institute, L'Aquila, Italy}
% ...
\affiliation[ff]{Nihon University Chiyoda, Tokyo, Japan} 
% ...
\affiliation[ld]{University of Alabama in Huntsville, Huntsville, AL, USA}
\affiliation[le]{Lehman College, City University of New York (CUNY), NY, USA}
% ...

\emailAdd{jeser@uchicago.edu}
\emailAdd{gfilippatos@mines.edu}
\emailAdd{thomas.paul@lehman.cuny.edu}
\emailAdd{guarino@na.infn.it}

\abstract{
        The complexity of modern cosmic ray observatories and the rich data sets they capture  often require a sophisticated software framework to support the simulation of physical processes, detector response, as well as reconstruction and analysis of real and simulated data.
        Here we present the EUSO-\Offline framework. The code base was originally developed by the Pierre Auger Collaboration, and portions of it have been adopted by other collaborations to suit their needs. We have extended this software to fulfill the requirements of Ultra-High Energy Cosmic Ray detectors and very high energy neutrino detectors developed for the Joint Exploratory Missions for an Extreme Universe Observatory (JEM-EUSO). These path-finder instruments constitute a program to chart the path to a future space-based mission like POEMMA. For completeness, we describe the overall structure of the framework developed by the Auger collaboration and continue with a description of the JEM-EUSO simulation and reconstruction capabilities. The framework is written predominantly in modern C++ (compliled against C++17) and incorporates third-party libraries chosen based on functionality and our best judgment regarding support and longevity. Modularity is a central notion in the framework design, a requirement for large collaborations in which many individuals contribute to a common code base and often want to compare different approaches to a given problem. For the same reason, the framework is designed to be highly configurable, which allows us to contend with a variety of JEM-EUSO missions and observation scenarios. We also discuss how we incorporate broad, industry-standard testing coverage which is necessary to ensure quality and maintainability of a relatively large code base, and the tools we employ to support a multitude of computing platforms and enable fast, reliable installation of external packages. Finally, we provide a few examples of simulation and reconstruction applications using EUSO-\Offline.
}


\arxivnumber{xxxx.yyyyy} 

\begin{document}
\maketitle
\flushbottom

\input{JemEUSO-OffLine_Introduction}
\input{JemEUSO-OffLine_Framework}

\input{JemEUSO-OffLine_DevOps}
\input{JemEUSO-OffLine_Sim_Reco}
\input{JemEUSO-OffLine_Conclusion}

\acknowledgments
First and foremost the authors would like to thank the Auger collaboration for providing the framework code as well as a majority of the simulation code. This allowed quick development and the confidence of using and adapting a well-tested software package.\\
This work was partially supported by Basic Science Interdisciplinary Research Projects of RIKEN and JSPS KAKENHI Grant (22340063, 23340081, and 24244042), by the ASI-INAF agreement n.2017-14-H.O, by the Italian Ministry of Foreign Affairs and International Cooperation, by the Italian Space Agency through the ASI INFN agreements Mini-EUSO n.2016-1-U.0, n. 2017-8-H.0, OBP (n. 2020-26-Hh.0) and n. 2021-8-HH.0, by NASA award 11-APRA-21730058, 16-APROBES16-0023, 17-APRA17-0066, NNX17AJ82G, NNX13AH54G, 80NSSC18K0246, 80NSSC18K0473, 80NSSC19K0626, and 80NSSC18K0464 in the USA, by the French space agency CNES, by the Deutsches Zentrum f\"ur Luft- und Raumfahrt, the Helmholtz Alliance for Astroparticle Physics funded by the Initiative and Networking Fund of the Helmholtz Association (Germany), by Deutsche Forschungsgemeinschaft (DFG, German Research Foundation) under Germany Excellence Strategy -EXC-2094-390783311, by National Science Centre in Poland grants 2017/27/B/ST9/02162 and 2020/37/B/ST9/01821, as well as VEGA grant agency project 2/0132/17, and by State Space Corporation ROSCOSMOS and the Interdisciplinary Scientific and Educational School of Moscow University "Fundamental and Applied Space Research".

\bibliographystyle{JHEP}
\bibliography{References.bib}

\end{document}

%% file: JemEUSO-OffLine_Introduction.tex
\section{Introduction}\label{sec:Intro}

The Joint Exploratory Missions for an Extreme Universe Observatory (JEM-EUSO) \cite{JEM-EUSO}
comprises a collection of experiments in pursuit of a future space-based cosmic
ray observatory, such as the Probe of Extreme Multi-Messenger Astrophysics
(POEMMA) \cite{POEMMA:2020ykm}. The objective of such an observatory is
elucidation of the origins and nature of Ultra-High-Energy ($E > 20~\rm{EeV}$)
Cosmic Rays (UHECR), and discovery of very high energy ($E >
20~\rm{PeV}$) neutrinos originating from astrophysical transient sources~\cite{transients}.

Several pathfinder missions designed to establish the technologies and
techniques to realize this objective have been completed or are in preparation.
EUSO-Balloon, a first prototype instrument flew on a short duration (one night)
high-altitude (\unit[40]{km}) balloon flight in 2014 \cite{2022SSRv..218....3A}. In 2017, a
long duration super pressure balloon flight was launched but gathered only
limited data owing to an apparent flaw in the balloon \cite{Wiencke:2017cfi}. A cross-calibration instrument,
EUSO-TA is deployed adjacent to a Telescope Array (TA) fluorescence station
\cite{Abdellaoui:2018rkw}. Since 2019, Mini-EUSO \cite{Bacholle:2020emk} has
been taking data on board the International Space Station. A second long duration super pressure balloon flight, EUSO-SPB2 was launched on May 13th from Wanaka, NZ~\cite{ICRC2021_EUSO-SPB2Overview, Eser:2023lck} but was prematurely terminated over the Pacific ocean due to a balloon leak after only 37 hours at float.

The first four of these pathfinders comprised telescopes capable of sensing the
fluorescence light produced by excited atmospheric nitrogen as the UHECR-induced
cascade of particles develops and deposits energy as it traverses the Earth's
atmosphere. These Fluorescence Telescopes (FT) all employed Fresnel lenses for
focusing, and increasingly refined versions of a Photo Detector Module (PDM) for
sensing the impinging photons. EUSO-SPB2 carried a science payload
comprising two telescopes. The first of these again employed the fluorescence
detection technique, but in this case, using Schmidt optics and 3 PDMs. The
second, a Cherenkov telescope (CT), was designed to detect the Cherenkov radiation emitted by the
particle cascade using silicon photomultipliers capable of ultra-fast integration
of the signal. This telescope can be pointed above or below the Earth's limb
to detect showers produced when cosmic rays or neutrinos respectively interact in the Earth producing up-going
particle cascades. While Cherenkov imaging has been utilized extensively on ground by gamma-ray telescopes such as MAGIC, HESS, or Veritas \cite{magic,HESS,VERITAS}, EUSO-SPB2 was the
first aerial mission to attempt this type of observation.

Simulating and reconstructing data from such a variety of instruments requires a
highly configurable and modular software framework. One software design with a
demonstrated history of success in this regard is the \Offline software developed by the Pierre Auger Collaboration starting in
2003~\cite{Argiro:2007qg}. In this article, we describe an extension of
the \Offline framework to accommodate the requirements of the JEM-EUSO
missions. An earlier (and complementary) software package, ESAF, is discussed
in~\cite{Berat:2009va, new_esaf} with a comparison between both frameworks outlined in the Appendix of \cite{new_esaf}.

This paper is organized as follows. In section~\ref{sec:Framework} we provide an 
overview of the framework design and discuss how the modularity and configuration requirements are satisfied.
Section~\ref{sec:DevOps} contains a description of the techniques used for swift
installation of external dependencies, environment virtualization, configuration 
and build, as well the methods used for testing coverage and continuous integration.
Section~\ref{sec:SimReco} contains a few examples of applications developed within
the EUSO-\Offline framework.  We offer some concluding remarks in section~\ref{sec:Conclusion}.

%% file: JemEUSO-OffLine_Framework.tex
\section{Framework}
\label{sec:Framework}

As noted in section~\ref{sec:Intro}, the EUSO-\Offline framework was inherited from the \Offline code project
initiated by the Pierre Auger Collaboration~\cite{Argiro:2007qg}, and
further developed by the NA61/SHINE software team~\cite{Sipos:2012hs} as well as some 
cosmic ray experiments using the radio detection technique~\cite{PierreAuger:2011btp}. The overall framework design has been largely retained for the EUSO-\Offline software. For the sake of completeness, we first  provide an overview of the framework design.

\subsection{Overview}

\begin{figure}[ht]
\centering
\includegraphics[width=0.5\textwidth]{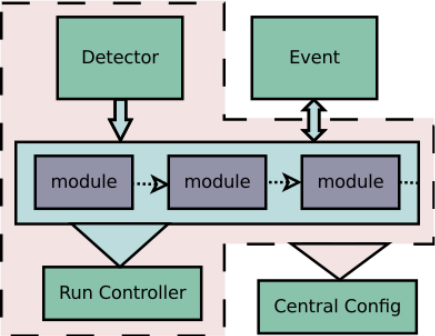}
\caption{General organization of the EUSO-\Offline framework. The blue-shaded 
region indicates interaction among modules, the Run Controller,
the Detector Description and the EDM. The pink-shaded region
indicates the components of the framework which are configurable via
the Central Config. See the text for detailed
explanation.}
\label{f:modules}
\end{figure}

The framework consists of the following principal components: 
a collection of event simulation and reconstruction algorithms contained in
{\em Modules}; a {\em Run Controller} which commands the modules to execute in a
particular sequence; an {\em Event Data Model} (EDM) from which modules read
information from other upstream modules and to which they write their results; 
a {\em Detector Description} which provides an interface to information about
the detector configuration as well as other data describing various conditions like atmospheric properties; and a {\em Central Config} which
directs the modules and framework components to their configuration data and
records provenance. The scheme of pipelining a collection of modules that communicate
via the EDM and read from the Detector Description separates
algorithms from data. This approach is not particularly Object Oriented,
but it does effectively support the collaborative development of simulation and reconstruction
algorithms. The framework components are illustrated in Fig.~\ref{f:modules}, and
discussed in more detail below.

\subsection{Modules and Run Control}

Simulation and reconstruction steps are generally organized in a simple pipeline.
Collaborators prepare algorithms in Modules.
 Modules inherit a common interface that declares the methods they 
carry out during processing, and which provides a macro to register
each module with the framework thereby making it aware of which 
modules exist for potential use at run time.

The modular design facilitates the comparison of algorithms and supports
building a variety of applications by combining modules in various
sequences. One can, for instance, swap out a module for reading in simulated
showers with a module to simulate laser shots (or other light sources) 
in the instrument field of view (FoV). Specification of a module sequence
is done at run time and does not require compiling any code. 

Module sequences are directed by the {\em Run Controller},
which invokes module execution according to a set of user-provided
instructions. A XML-based tool is used for specifying sequencing
instructions, including support for nested loops and loops which 
iteratively process the data. Modules can signal the Run Controller
to break a loop or skip downstream modules if certain conditions arise.
For instance, if one step in a collection of reconstruction algorithms
fails, a signal to skip to the next event can be relayed to the Run
Controller.

The approach of using simple instructions provided in an XML file
together with a mechanism for modules to signal the Run Controller
has proved sufficiently flexible for our applications. However, 
we are planning to introduce Python bindings which will allow a 
more flexible system in the future.

\subsection{Event Data Model and Detector Description}

The \Offline framework provides parallel hierarchies for accessing data. The
{\em Event Data Model} is used for reading and writing information specific for each event. The read-only 
{\em Detector Description} supports retrieval of static configuration
data or relatively slowly varying information such as detector health or 
atmospheric monitoring data.

\subsubsection{Detector Description}
\label{sec:DetectorDescription}
Configuration and conditions information accessible via the Detector Description 
can include detector materials and geometry as well as time-dependent information such
calibration data, background measurements, and atmospheric monitoring data. 
The Detector description is meant to provide a single endpoint for these sorts
of data, preventing possible errors resulting from inadvertently 
providing the same information in different pieces of code or configuration
files. 

\begin{figure}[ht]
\centering
\includegraphics[width=0.99\textwidth]{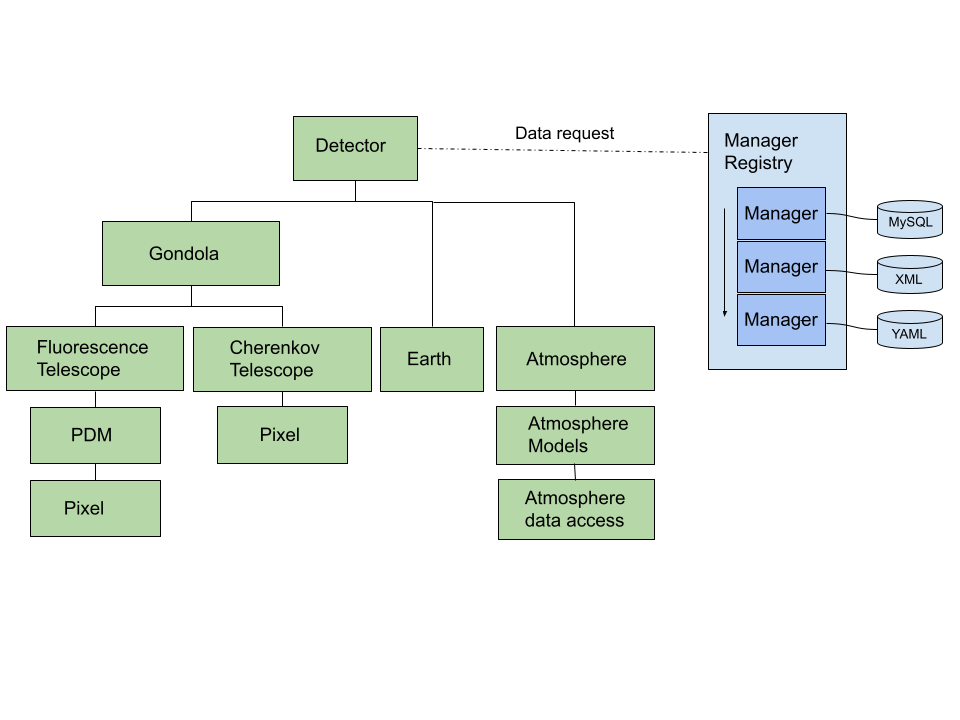}
\caption{Manager mechanism for the case of the EUSO-SPB2 instrument. Left: The User Interface to detector information
comprises a hierarchy of objects that 
follow the hierarchy naturally associated with the Detector. Modules can query this interface 
for information needed during data processing. The science payload lives
under the Gondola, and users can query for things like the Gondola orientation,
or the configuration of the FT and CT instruments that live below it in the 
hierarchy. Notice that, the Atmosphere and the Earth are also
considered to be parts of the detector (since the Atmosphere functions as a 
giant calorimeter). Notice that the Detector interface 
in this figure is structured to follow the hierarchy associated with the SPB2 instrument. The structure can
be selected using the \Offline configuration mechanism. For example, one can select the EUSO-TA instrument 
configuration in an XML file, and an interface will be constructed which is appropriate to that instrument (that is, with no Cherenkov
telescope and no gondola objects.) An additional mechanism is also available 
to perform pre-processing of data read by various Managers. We use this
primarily to provide higher-level access to atmospheric data. Various {\em Atmosphere Models} can be accessed to compute
quantities such as fluorescence yield, scattering, and absorption. 
The Earth object can be queried for albedo estimates for the sea or different
sorts of terrain.
Right: The Detector sends its requests to the Manager Registry, 
which finds the appropriate data source and returns the requested data to the Detector
interface. See the text for more detail.
}
\label{f:managers}
\end{figure}

The Detector Description interface is structured to follow the hierarchy
normally associated with actual detector components. Requests sent to 
the Detector Description are relayed to a registry of {\em Managers}, each
of which is capable of extracting particular data from a particular source.
In this way, the user sees a single interface, while a back-end of Managers
deals with the potentially involved task of reading data from different sources.
For example, different Managers might read data from XML files or from 
databases, thus isolating the (possible) complexity of data access in
manageable units of code.  The general scheme is illustrated in Fig.~\ref{f:managers}.

Managers in the registry are polled for information using a chain
of responsibility, as indicated by the arrow in Fig. \ref{f:managers}.
When the registry receives a request, it asks the first manager in the 
registry if it can answer the request. If it cannot, it proceeds to 
the next manager in the chain. This allows for fall-back solutions
to common problems. For instance, time-dependent information stored
in a database may have missing entries for some time intervals. 
This can be addressed by providing a fall-back manager downstream from
the manager that provides database access. Specification of 
which data sources are accessed and in what order are provided
in a configuration file. 

A plug-in mechanism in the Atmosphere supports {\em Atmosphere Models}.
These models are used to compute fluorescence and Cherenkov yields, Rayleigh and Mie
scattering and absorption as well as ozone absorption, as further
described in Section~\ref{subsec:Simulations}. These models can be 
configured to access parametric data stored in configuration files or
time-dependent data stored in databases. A template is provided to assist
collaborators in developing different interchangeable Atmosphere Models, in much the same 
vein as interchangeable simulation and reconstruction Modules can be developed.
Model selection is afforded at run time
through a configuration file read by a model registry which works in 
the same manner as the manager registry discussed above. More details on this plug-in mechanism can
be found in reference~\cite{Argiro:2007qg}.

Lazy evaluation is used to access all detector data so that only requested
data are actually loaded into the Detector Description and cached for
future access, possibly flagged with a time interval for which the data are valid.
Data that are valid only for a specific time interval
are flushed and reloaded when an event is read which 
contains a time stamp outside the interval of validity. As the Detector
is meant to provide a read-only interface, the entire interface is 
logically \textit{const}, though it is physically non-\textit{const} since it must load itself
from the various data sources.

\subsubsection{Event Data Model}
\label{subsection:EDM}

The {\em Event Data Model} (EDM) contains the raw, calibrated and 
reconstructed data as well as Monte Carlo true parameters for the case of simulations.
It provides the backbone for communication among physics modules.
The structure of the EDM comprises a collection of classes organized
with the hierarchy normally associated with the detector, similar 
to the Detector Description interface. Physics modules access the 
event information through a reference to the top of the hierarchy, which
is provided to the module interface by the Run Controller. Since the 
Event enables inter-module communication, reference semantics are used
to access objects in the EDM, and constructors are private in order 
to prevent accidental copying.

The EDM is instrumented with a protocol allowing modules to discover its
constituents at any point in processing, and thereby determine whether the input
data required to carry out the desired processing are available (and take
appropriate action if not). The event interface cannot be modified by 
Modules. This somewhat limits flexibility, but it does help to ensure
interchangeability of physics modules, an essential feature
of \Offline as it allows collaborators to try out different approaches
to the same problem without interfering with one another.

\Offline is also equipped to populate parts of the event by reading formats employed
by most popular air shower simulation packages~\cite{Heck:1998vt, Bergmann:2006yz, Sciutto:1999jh, PhysRevD.103.043017},
as well as data formats used by the different JEM-EUSO pathfinders. 
A {\em DataWriter} module is available to write out the event in the same
ROOT formats employed by the JEM-EUSO instruments; simulation true parameters is stored
in a separate ROOT tree in the same file(s) containing reconstruction and experimental configuration information.

\subsection{Configuration}

\Offline provides an XML and XML Schema based system to
organize data used to configure the software for the variety of simulation
and reconstruction tasks required to support the various JEM-EUSO instruments.
The \emph{Central Config} configuration tool points modules and framework components
to the location of their configuration data and connects to
Xerces-based~\cite{xerces} XML parsers to assist in reading information from
these locations. The Xerces API is wrapped with a custom interface which provides
a simpler interface at the cost of somewhat reduced flexibility.
Our XML parser also provides
conveniences such as automatic casting to JEM-EUSO data containers as well as automatic unit
conversion, obviating the need to enforce a prescribed units convention.
Syntax and content checking of the configuration files are implemented using W3C
XML Schema validation. Xerces was adopted early on by the Pierre Auger
Observatory \Offline~\cite{Argiro:2007qg} partly because of its comparatively complete 
support for the XML Schema standard.

The {\em Central Config} records {\em all} configuration data accessed during a
run and stores them in an XML log file. The log file can subsequently be read in to
reproduce a run with an identical configuration. This allows collaborators to
exchange configuration data and compare results. The logging mechanism is
also used to record the external libraries which are
used for each run.
We have found such provenance tracking to be indispensable for simulation
production and data analysis; years of experience have shown that it always happens that someone eventually
needs to look up precisely what configuration was used to produce a simulation or 
process data.

%% file: JemEUSO-OffLine_DevOps.tex
\section{DevOps}
\label{sec:DevOps}
In this section, we discuss the tools we have adopted for code distribution,
continuous integration and deployment (CI/CD), code building, and dependency
handling.

The EUSO-\Offline code repository is hosted on GitLab~\cite{gitlab}. In addition to
repository hosting, we employ most of the GitLab project management tools. 
Code development is conducted on dedicated branches documented in merge 
requests and peer-reviewed before merging into the main branch. Issues are tracked using 
the GitLab issue tracking system, which allows for tight integration of issues and Merge Requests (MRs).

A significant effort has been devoted to establishing thorough
testing coverage. This is crucial for all software projects
of any size, and for the case of JEM-EUSO the need is particularly pronounced
owing to the multiple experimental configurations which must be supported.
The GitLab CI/CD is used to build the code using both the gcc and clang
compilers with each check-in, as well as to run unit tests, regression tests, linting and static analysis using Clang-Tidy~\cite{clang-tidy} and Clang Static Analyzer~\cite{clang-static-analyzer}. 
The built-in clang sanitizers are employed to detect
run-time errors. Older unit tests are implemented with the help of CppUnit~\cite{cppunit}, while
more recently implemented tests are based on GoogleTest~\cite{gtest}.
Regression tests run full sequences of modules to detect unexpected changes in simulation and reconstruction results.
We have developed an in-house tool to assist with this. The build system employs CMake~\cite{cmake} to write the build tool configuration (for example GNU Make~\cite{make} or Ninja~\cite{ninja}).

The external packages upon which EUSO-\Offline is built were to some extent
inherited from the Pierre Auger \Offline software. The choices of externals were
dictated not only by functionality and open-source requirements, but
by the best guess at longevity. Some functionality is provided to 
the client code via fa\c{c}ads, as in the case of reading XML files, or
via a bridge, as in the case of the detector description described previously. The collection of external libraries includes
Xerces~\cite{xerces} for XML parsing, CLHEP~\cite{clhep} for expressions evaluation, Geant4~\cite{Agostinelli:2002hh, Allison:2006ve, Abdellaoui:2018rkw} for detailed detector
simulation, Boost~\cite{boost} for its numerous C++ extensions,
mysql~\cite{mysql} and sqlite~\cite{sqlite} for database implementations, pytorch-cpu~\cite{pytorch} for machine learning
applications and ROOT~\cite{root} for file input and output.
The use of ROOT has been limited to input/output since the first design of \Offline in the early 2000s owing to issues at the time with bugs, the non-idiomatic C++ design, and concerns about vendor lock-in; more recently the extensive collection of
readily available solutions provided by the huge open-source community
mostly obviates the need for ROOT utilities, with the arguable exception of input/output.

We use Anaconda (a.k.a conda)~\cite{conda} together with the Libmamba~\cite{mamba} dependency solver to install all externals as pre-compiled binaries and to create a virtual environment largely
isolated from the local system, which in turn simplifies the package detection performed by CMake.
This allows for dependency installation that takes a few minutes rather than the many hours typically required to build large packages like Geant4 and ROOT from source. We also generate conda packages of in-house software like EtoS and EtoT~\cite{Piotrowski:2014tsa}. 

%% file: JemEUSO-OffLine_Sim_Reco.tex
\section{Simulation and Reconstruction Capabilities} \label{sec:SimReco}

In this section, we provide an overview of some of the simulation and
reconstruction capabilities of the EUSO-\Offline 
framework. One  purpose of this section is to
illustrate some of the advantages of algorithm modularization and flexible
configuration. We also demonstrate some of the benefits
of implementing simulation, reconstruction and
analysis of all the JEM-EUSO missions in a single, common framework. 

The majority of the simulation modules were originally developed by members of the Pierre Auger collaboration, starting with input of shower simulation in any of several popular shower generator output formats and continuing with the light production and the transport of photons through the atmosphere \cite{AugerFdSim}. For the JEM-EUSO missions, it was necessary to extend the code in a number of ways, itemized below:

\begin{itemize}
\item The effects of Ozone absorption is included as a {\em model}
  accessible via the Atmosphere interface.
  \vspace{-0.2cm}
\item The Earth interface is added to the DD to provide access to
  albedo estimates of different sorts of terrain which can be used
  for simulation of Cherenkov light reflected from the Earth's surface
  into an orbiting telescope.
  \vspace{-0.2cm}
\item A reader for the EASCherSim~\cite{PhysRevD.103.043017,
  PhysRevD.104.063029} Monte Carlo generator is available.
  This generator supports the simulation of Cherenkov light emitted at very
  small angles with respect to the shower axis.
  \vspace{-0.2cm}
\item A configurable Geant4-based telescope simulation module
  is available, which can model any of the JEM-EUSO instrument designs
  by simply specifying the desired XML configuration.
	\vspace{-0.2cm}
\item Custom Fresnel optics simulation (written originally for ESAF \cite{ESAF-lens}) is incorporated into the Geant4 simulation of the telescopes.
	\vspace{-0.2cm}
\item Background simulation modules (both night-glow and spot-like) is
  prepared and can model background light observed by the Fluorescence
  telescopes in either tilt or nadir pointing modes.
	\vspace{-0.2cm}
\item Simulation of the trigger logic for the different instruments has been
  prepared.
  \vspace{-0.2cm}
\item  A convolutional neural network was developed and trained on simulated data
  to perform fast in-flight classification of events in order to prioritize
  event downlinking~\cite{Filippatos_2021}.
\end{itemize}

\subsection{Simulation}
\label{subsec:Simulations}
Simulations generally commence either with input from a cosmic ray air shower
generator, or simulation of calibration laser shot.  A common input/output
interface, discussed in some detail in~\cite{Argiro:2007qg}, connects the \Offline
Event to the appropriate back-end reader. We currently provide readers
for the air shower generators \texttt{CORSIKA}~\cite{Heck:1998vt},
\texttt{CONEX}~\cite{Bergmann:2006yz} and \texttt{EASCherSim}~\footnote{\url{https://c4341.gitlab.io/easchersim/index.html}}. Laser simulation is performed by a module distributed
with the \Offline software.  Simulation of fluorescence detection is then
performed by a sequence of Modules under the direction of the Run Controller and
sequencing instructions provided in a configuration file, as explained in
section~\ref{sec:Framework} and outlined in more detail below.

The first module in the FT simulation sequence positions the generator-level
shower either by defining its distance and azimuthal orientation with respect to
the telescope's optical axis (for nearly horizontal showers observed by
balloon-borne instruments) or by pinning the shower axis to a position on the
Earth (for the case of vertical showers, as measured for instance by
EUSO-TA). At this step, it is possible to randomize the positioning parameters and
to impose cuts based on the instrument field of view (FoV). FoV cuts increase the speed of simulations
with randomized positioning parameters, since subsequent modules can then deal
only with visible portions of the shower.

Once the shower is positioned, a downstream module converts the charged particle
distributions provided by the shower generators \texttt{CORSIKA} or \texttt{CONEX} into
fluorescence and Cherenkov photons. These conversions are performed with the
help of fluorescence yield and Cherenkov generation Atmosphere Models using
the configurable back-end of the Atmosphere interface described in
section~\ref{sec:DetectorDescription}. A number of different models are available
and selectable at run time using a configuration file. At present, we estimate
the Cherenkov emission using a parametric approach following Hillas or
Nerling~\cite{Hillas_1982, Nerling_2006}.  The fluorescence yield is estimated
using the approach provided in Auger-\Offline, in which
wavelength-dependent measurements from Airfly~\cite{Ave:2007xh} are combined
with the Nagano absolute fluorescence yield estimate~\cite{Nagano_2004}.
Ozone absorption is computed using data from the ozone sounding program of 
the US National Oceanic and Atmospheric Administration (NOAA)~\cite{noaa}. Details 
are provided in~\cite{Falk:2015voa}.

Subsequent modules simulate the light propagation from the shower to the
detector aperture. Again, several configurable Atmosphere Models are available
to assist in the computation of the effects of Rayleigh and Mie scattering and
absorption as well as ozone absorption, the latter of which is particularly
consequential for a space-based observatory \texttt{EASCherSim} contains its
own Rayleigh, Mie and ozone models. At this point, the most computationally
intensive parts of the simulation are complete and results can be written to
file by the {\em DataWriter} described in section~\ref{subsection:EDM}.
Using these partially simulated events, we can more quickly develop
downstream detector simulation and triggering algorithms.

\begin{figure}[tb]
 	\centering
 	\begin{subfigure}[a]{0.59\textwidth}
     	\centering
     	\includegraphics[width=\textwidth]{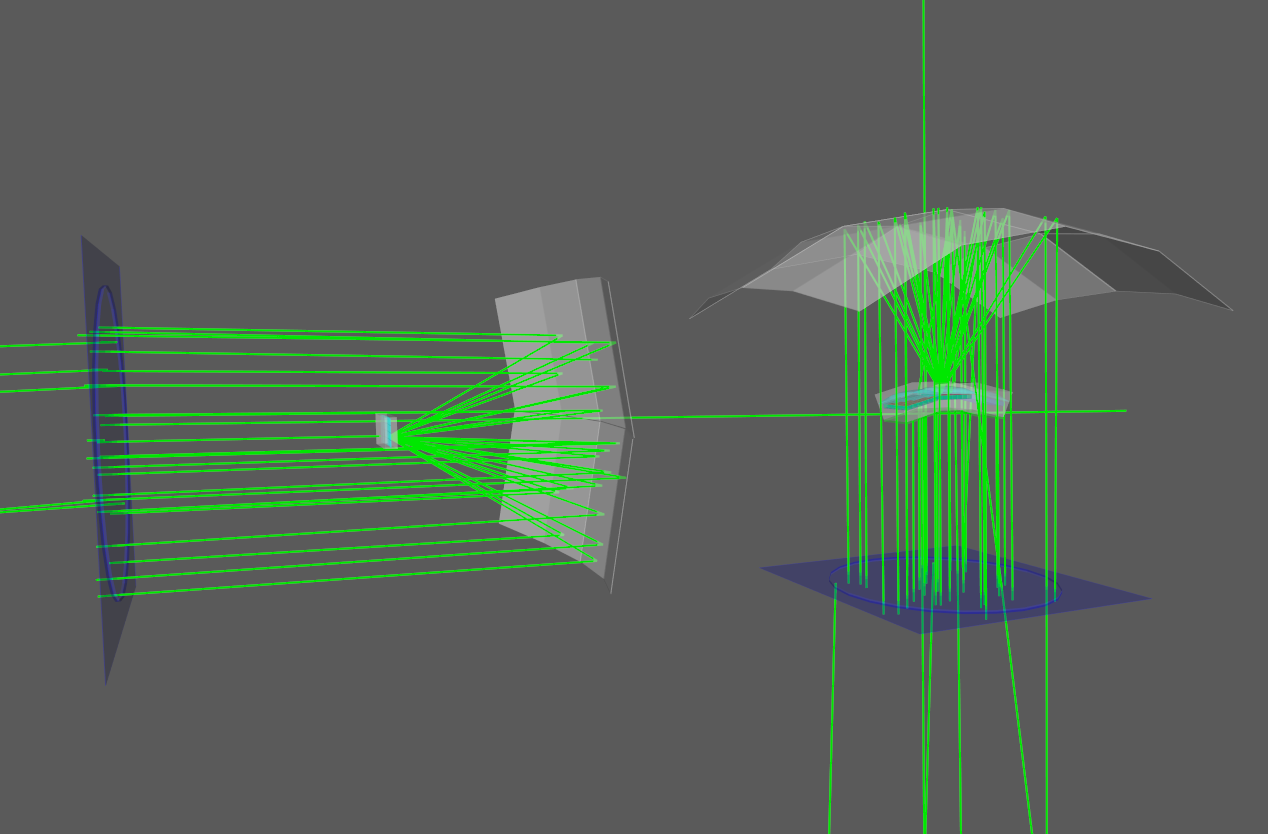}
     	\caption{
       	Geant4 simulation of the EUSO-SPB2 instrument, which employs Schmitt optics. The green lines
       	represent photons; for clarity of the image, the number of photons
       	injected is far fewer than in a typical event. Mirrors and cameras are
       	shown in grey, and the blue regions represent the entrance pupils.
       	The instrument on the right is the Fluorescence telescope, which focuses light
        on 3 Photo Detection Modules. The Cherenkov
       	telescope is on the left, and employs silicon photomultipliers. The Cherenkov telescope can be rotated to view the
       	regions just above and below the Earth's limb.
     	}
 	\end{subfigure}
 	\hfill
 	\begin{subfigure}[a]{0.38\textwidth}
     	\includegraphics[width=\textwidth]{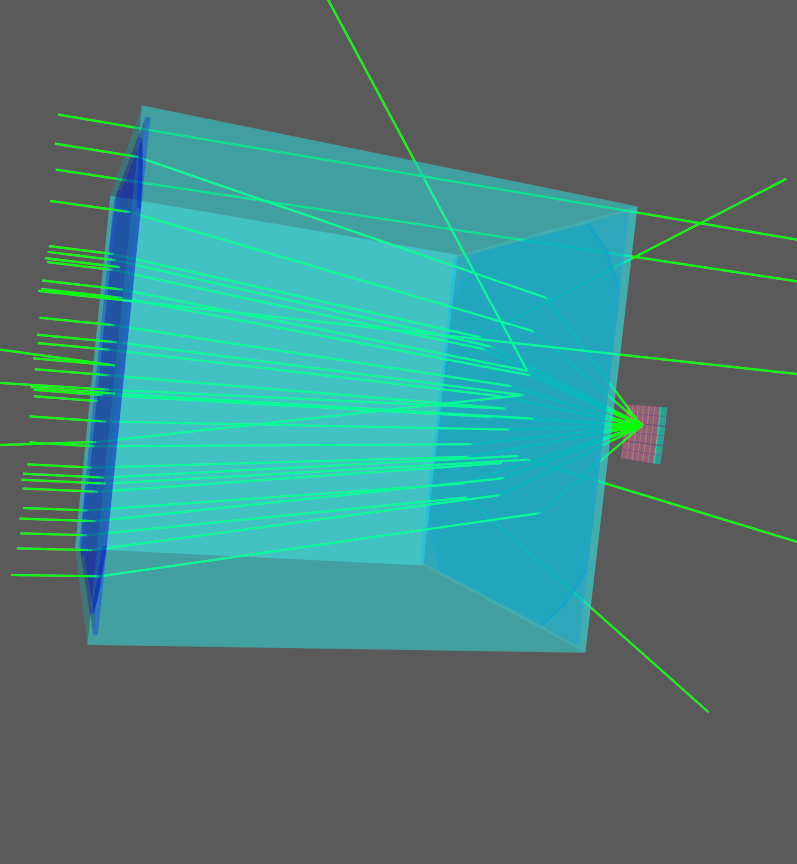}
     	\caption{
       	Geant4 simulation of the EUSO-TA instrument, which comprises 2 Fresnel
       	lenses and 1 Photo Detection Module.
     	}
 	\end{subfigure}
  \caption{Geant4 simulations of the EUSO-SPB2 (a) and EUSO-TA (b) instruments.}
\label{fig:Geant4SPB2example}
\end{figure}

Once the number of photons arriving at the detector aperture has been
computed, full detector simulation can be performed. Our method-of-choice
for simulating light propagation from the aperture to the sensors employs
Geant4 for ray tracing through the various optical interfaces. Simulated event examples of the EUSO-SPB2 payload and the EUSO-TA instrument 
are shown in Fig.~\ref{fig:Geant4SPB2example}.
The Geant4 simulation can be configured to model all of the JEM-EUSO missions.

Simulation of the camera efficiency, electronics response, and digitization is
performed by our own custom modules which are tuned using laboratory
measurements.

Downstream modules perform background simulations.  Night sky background is
accounted for by appending the signal trace of each pixel with a background
component derived from measurements, including the effects of tilting the
optical axis. Spot-like events from terrestrial light sources are simulated
using data recorded by different JEM-EUSO instruments. The pre-trigger 
fully simulated event can be written to file at this stage.

\begin{figure}
     \centering
     \begin{subfigure}[b]{0.49\textwidth}
         \centering
         \includegraphics[width=\textwidth]{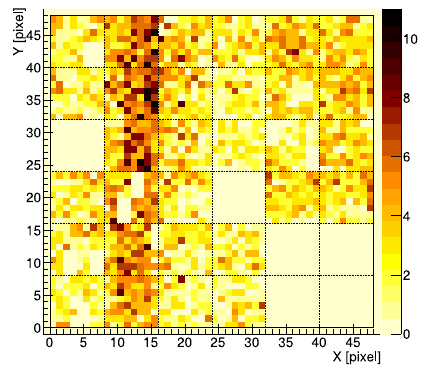}
         \caption{Data}
     \end{subfigure}
     \hfill
     \begin{subfigure}[b]{0.49\textwidth}
         \centering
         \includegraphics[width=\textwidth]{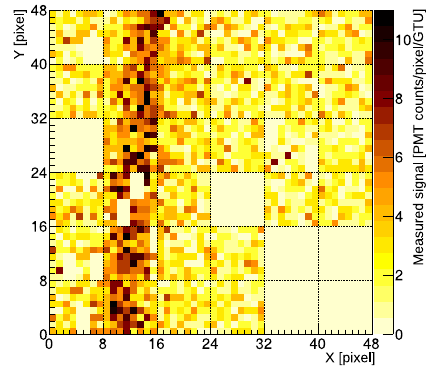}
         \caption{Simulation}
     \end{subfigure}
    \caption{A UHECR track with an energy of
      \unit[10$^{18}$]{eV} and an impact point \unit[2.6]{km} from the
      detector. The color code shows the counts per pixel per time frame of \unit[2.5]{$\mu$s}(Gate Time Unit). The shower parameters were provided to us by the TA collaboration based on their reconstruction of this shower recorded at the TA site Black Rock Mesa. The zenith angle of the shower was 8\textdegree~ and its azimuth
      was 82\textdegree. Panel (a) shows the track as recorded in EUSO-TA while
      panel (b) shows the simulation of such a shower within EUSO-\Offline. Figure from \cite{Abdellaoui:2018rkw}.}
    \label{fig:DataSimComp}
\end{figure}

Triggering algorithms are next applied to the background-contaminated signal.
This step takes advantage of the modular design, as different trigger modules
can be used together or separately based on the instrument in question and the
algorithm under evaluation. Triggered events can be written to file in the same
format as the JEM-EUSO mission data, or they can be passed directly to
subsequent reconstruction modules, discussed in
section~\ref{subsec:Reconstruction}. Fig.~\ref{fig:DataSimComp} shows a
comparison between a shower recorded
by EUSO-TA \cite{Abdellaoui:2018rkw} and a shower of (approximately) equivalent
energy and geometry simulated with EUSO-\Offline. Parameters for the
shower simulation are based on reconstruction results provided by the Telescope
Array collaboration~\cite{TAover}.

\subsection{Reconstruction}
\label{subsec:Reconstruction}

The structure of EUSO-\Offline enables straightforward validation of reconstruction
processes using simulated data since simulations record ground-truth in the
Event Data Model. For example, photons in the simulation keep track of
whether they originate from fluorescence, Cherenkov, or laser,
so the photons arriving in each pixel can be broken into components according to their
origin. The simulation also keeps track of whether a photoelectron was produced due to a signal of interest or background. As mentioned above, reconstruction can be performed in the same run as
simulation or can commence from libraries of fully (or even partially)
simulated events.  Reconstruction of simulated and real data can be naturally
broken up into a pipeline of processing modules, each of which can be approached
using different algorithms which can be compared to one another using the
mechanisms discussed in Section~\ref{sec:Framework}. We outline here the modules currently in use. In the final sub-section we briefly discuss
using machine learning inside the EUSO-\Offline framework.

\begin{figure}
  \includegraphics[width=\textwidth]{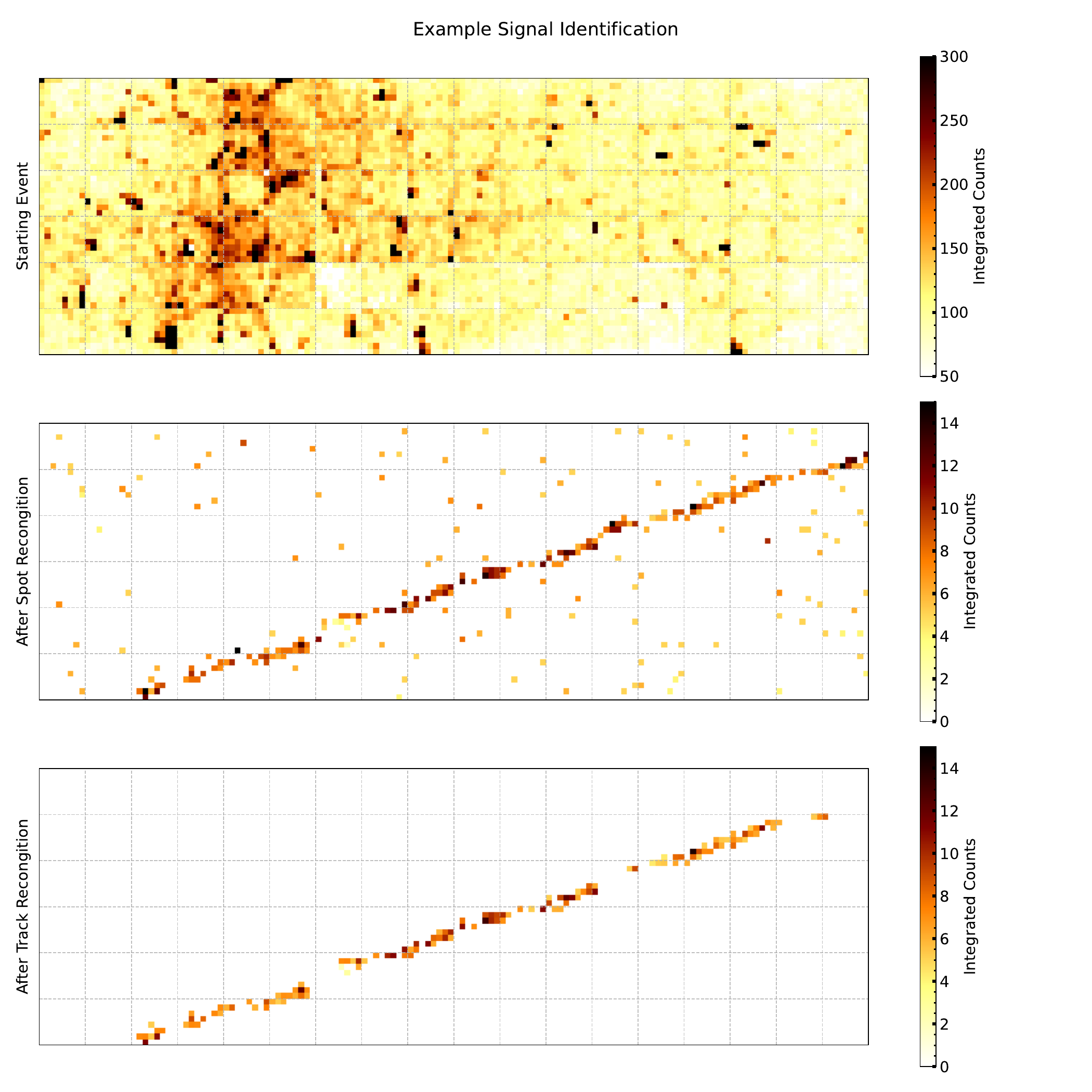}
  \caption{Recorded laser track on the EUSO-SPB2 focal surface during the field campaign in the summer of 2022 at the TA site. The x and y axis represent the camera x and y coordinates in units of pixels. Telescope pointing at nadir, laser located 24 km away travelling nearly horizontally. Integrated counts over 128 frames (top), pixels remaining after spot identification (middle), and pixels remaining after track identification (bottom).}
\label{fig:BackgroundSubtraction} 
\end{figure}

The goal of the reconstruction is to identify the main properties of the incoming air shower, which are arrival direction, primary energy, and composition. But before this can be done, the pixels containing the light of the observed air shower (or laser) need to be identified and the backgrounds need to be removed, both for simulated and real data. An example of progression from noisy data to identified track is shown in Fig. \ref{fig:BackgroundSubtraction}.

The background removal is performed on a pixel-by-pixel basis requiring a threshold for the signal based on the mean and standard deviation of the individual pixel trace. By default a 5 $\sigma$ excess is required but is a configurable parameter. The background removal process is independent of the trigger conditions. Depending on the parameters used, an event which passes the hardware trigger, or simulated trigger, may be entirely or partially removed by the background subtraction.

In order to identify a track within the remaining pixels, the subsequent modules clean the track by requiring those pixels containing the track not to be isolated. Only pixels that are part of a cluster of at least N pixels within a given radius are kept. In addition, only clusters that have neighboring clusters in time are considered to guarantee a moving track in the camera as expected for an air shower signal. An example of such a ``cleaned'' track is shown in the middle of Fig \ref{fig:BackgroundSubtraction}. Finally, a line is fitted to the ``cleaned'' track and all remaining outlier pixels are removed.

Subsequent modules determine the properties of the primary particle that created the identified track in the data. The first step is a geometrical reconstruction deploying a well-established standard method. The Shower Detection Plane (SDP) is defined using the pointing of each signal pixel weighted by its calibrated signal strength. The defined SDP allows one to determine an elevation angle for each signal pixel. These angles can be used in the next module to determine the shower direction based on a time-angle fit. More details on this method can be found for example in \cite{JEM-EUSO:2018tmw}. The technique is also used by the Pierre Auger and TA collaboration for their geometric reconstruction.

With the geometry of the shower in hand, a final module uses the number of photons detected to estimate the energy of the shower (or calibration laser shot). This is done by summing all the signals recorded at the detector and assuming isotropic emission from the source, which is optimal for lasers. For the case of an EAS, the data are fitted to a Gaiser Hillas function, the maximum of which provides the shower $X_{\rm max}$ parameter. The total energy deposited by the shower is determined by integrating the fitted function~\cite{hillas_1972}.

\subsection{Machine Learning Methods}
\label{subsec:MLMethods}
Using the EUSO-\Offline machinery, we have developed a convolutional neural network for
in-flight event classification during the EUSO-SPB2 \cite{Filippatos_2021}
mission. This neural network is trained on simulated data prior to flight and
runs onboard the instrument using raw data.

EUSO-\Offline is currently distributed with the torch libraries~\cite{NEURIPS2019_9015}. By utilizing Just
In Time (JIT) compilation tracing, models trained in a variety of environments can be
loaded into EUSO-\Offline for inference. This added level of flexibility allows for
easy testing of models on a large variety of data, without the added overhead
that is needed for training complex models, such as GPU-accelerated computation.
JIT compilation also allows for different architectures to be applied to data and simulations, allowing for EUSO-\Offline to utilize the latest available machine learning methods without changes to the framework. 

The modular nature of \Offline allows for a greater deal of flexibility in utilizing machine learning techniques. For example, the onboard classification scheme for EUSO-SPB2, which utilized a recurrent convolutional neural network, exists as a module which can be added to any sequence, at any point. This allows for various models to be applied to simulated data quickly, which was useful in the development of the in-flight classification method. The binary classifier can also be applied to recorded data in parallel to the traditional track finding methods described in Section~\ref{subsec:Reconstruction}, and the performance of the two methods can be directly compared.

While the current use cases of machine learning in EUSO-\Offline have been focused on
binary classification, the machinery exists for more complex analysis. We
anticipate applying machine learning methods to address some of the
reconstruction tasks discussed above in Section~\ref{subsec:Reconstruction}.

%% file: JemEUSO-OffLine_Conclusion.tex
\section{Conclusion}\label{sec:Conclusion}
We have adapted the \Offline framework originally developed by the Pierre Auger Observatory to the requirements of 
the JEM-EUSO missions. This framework provides the essential machinery to facilitate collaborative development of algorithms
to address the variety of simulation and reconstruction use-cases required to analyze data from current and pending JEM-EUSO mission configurations. The modular design provides a straightforward way to compare different approaches to a given problem, and the configuration machinery provides the flexibility necessary to deal with the variety of simulation and reconstruction applications as well as the different instrument designs. 
The user interface to event data and time-dependent detector and atmospheric data is kept as simple as is feasible, while the back-end handles the complexity required to deal with different data sources and formats. This software has been used in production for the analysis of data from the EUSO-Balloon, EUSO-TA, EUSO-SPB1 and EUSO-SPB2 missions.\\
While at the moment the software package is available to members of the JEM-EUSO collaboration, a strategy is being developed to follow the OpenData/OpenSource \cite{Berlin} principles. This requires a close communication with the Pierre Auger collaboration which is proprietary of the common, base part of the code. Its current policy requires the acceptance of the Pierre Auger Collaboration before using the code. In the mean time other avenues like publishing only independent parts of the package are being investigated. Furthermore the outlined practice in Section \ref{sec:DevOps}, following the industry standard for software development, provide the basis for a stable, maintainable code base. Once the code is public, its design principle of modularity and high configurability as well as a automatic generated API documentation directly guarantees that \Offline follows the FAIR \cite{FAIR} principles.